\documentclass[preprint,12pt,1p]{elsarticle}

\usepackage{amssymb}
\usepackage{amsthm}
\usepackage{amsmath} 
\usepackage{color}

\journal{Physics Letters A}

\begin{document}

\begin{frontmatter}



\title{Ordinal language of antipersistent binary walks}


\author[inst1]{Felipe Olivares}

\affiliation[inst1]{organization={Instituto de Fisica Intersciplinar y Sistemas Complejos IFISC (CSIC-UIB)},
            city={Palma},
            postcode={07122}, 
            country={Spain.}}

\begin{abstract}
This paper explores the effectiveness of using ordinal pattern probabilities to evaluate antipersistency in the sign decomposition of long-range anti-correlated Gaussian fluctuations.  It is numerically shown that ordinal patterns are able to effectively measure both persistent and antipersistent dynamics by analyzing the sign decomposition derived from fractional Gaussian noise. These findings are crucial given that traditional methods such as Detrended Fluctuation Analysis are unsuccessful in detecting anti-correlations in such sequences. The numerical results are supported by physiological and environmental data, illustrating its applicability in real-world situations.
\end{abstract}




\end{frontmatter}



\section{Introduction}
Long-range correlations with scale-invariant structures are present in a wide variety of natural phenomena. These macro-scale properties are normally captured by temporal fluctuations of representative variables that describe certain properties of the micro-scale components driving the dynamics of the system under study~\cite{artime2024robustness}. It has been found that those fluctuations can be decomposed into their magnitude (absolute value) and their sign (direction), where each of them carries different information of the dynamics~\cite{ashkenazy2001magnitude}. More specifically, the former accounts for non-linear correlations and multifractal properties, while the sign decomposition relates to the linear ones. This approach has been widely used in several areas such as the heart dynamics~\cite{ashkenazy2001magnitude,kantelhardt2002characterization}, finance~\cite{liu1999statistical}, fluid dynamics~\cite{zhai2017nonlinear}, terrestrial temperature~\cite{bartos2006nonlinear} and ocean temperature~\cite{kalisky2005volatility}. This decomposition procedure greatly facilitates the modeling surrogate time series and helps to reveal whether there is a coupling in the mechanisms responsible for both the magnitude and direction of the data values~\cite{gomez2016magnitude}. 

The Hurst exponent is commonly used to quantify scaling laws for long-range correlations. One popular approach to estimate it is through Detrended Fluctuation Analysis (DFA)~\cite{kantelhardt2001detecting}, especially for real-world fluctuations. The Hurst parameter, $H$, defines two distinct regions in the interval $(0,1)$, separated by $H=1/2$ that corresponds to completely uncorrelated fluctuations (white noise). When $H>1/2$, consecutive values tend to have the same sign so that these processes are persistent.  Conversely, for $H<1/2$, consecutive values are more likely to have opposite signs,  indicating antipersistent dynamics~\cite{feder2013fractals}.  It has been analytically and numerically shown that for $H\geq1/2$, the sign decomposition of fractional Gaussian noises (fGn) inherits the linear correlations  with its exponent $H_{\text{sign}}$ asymptotically approaching the same value as $H$ that characterize the original fluctuations~\cite{gomez2016magnitude}. On he other hand, for $H<1/2$, despite the antipersistent fluctuations, its sign series is uncorrelated at large temporal scales---$H_{\text{sign}}\approx1/2$~\cite{gomez2016magnitude}. 

Carpena and co-workers have shown analytical and numerical evidence that this supposed impossibility of generating anti-correlations in binary series is, in fact, a spurious result when DFA is applied to the sign series obtained from fGn, for estimating the Hurst exponent~\cite{carpena2017spurious}. Actually, they have shown that anti-correlations in the sign sequences can be effectively  quantified using the autocorrelation function~\cite{carpena2017spurious}, but it requires long time series ($\approx 2^{24}$ data points), a condition rarely met in real-world experiments. Additionally, using the autocorrelation function to quantify long-range correlations is often impractical because it tends to be noisy and highly sensitive to the size of the time series~\cite{coronado2005size}. Previous studies have shown that even for series lengths of up to $2^{20}$ data points, there is a tendency to overestimate the correlation exponent~\cite{coronado2005size}. This underscores the challenge of developing new methodologies for assessing anti-correlations in finite binary sequences.

This work addresses the characterization of anti-correlations of binary sequences obtained from the sign decomposition of fractional Gaussian noise, using ordinal patterns~\cite{bandt2002permutation}. One advantage of using this symbolization technique is that the patterns naturally emerge from the time series without requiring any model-based assumptions. Additionally, this method only requires a very weak stationary assumption~\footnote{Being $D$ the pattern length, the weak stationary assumption is, for $k = D$, the probability for $x_t < x_{t+k}$ should not depend on $t$~\cite{bandt2002permutation}}, unlike traditional autocorrelation analysis, which is applicable only to stationary time series~\cite{kantelhardt2001detecting}. The goal of this study is to provide a numerical and empirical benchmark for analyzing (anti-)correlated binary time series. The use of ordinal patterns probability can effectively evaluate both persistent and antipersistent dynamics in the sign decomposition. The study also compares the numerical results with empirical anti-correlated fluctuations found in physiological and environmental data, highlighting its practical utility.

\section{Ordinal patterns probability}

\begin{figure}[t!]
\centering
\includegraphics[width=0.8\textwidth]{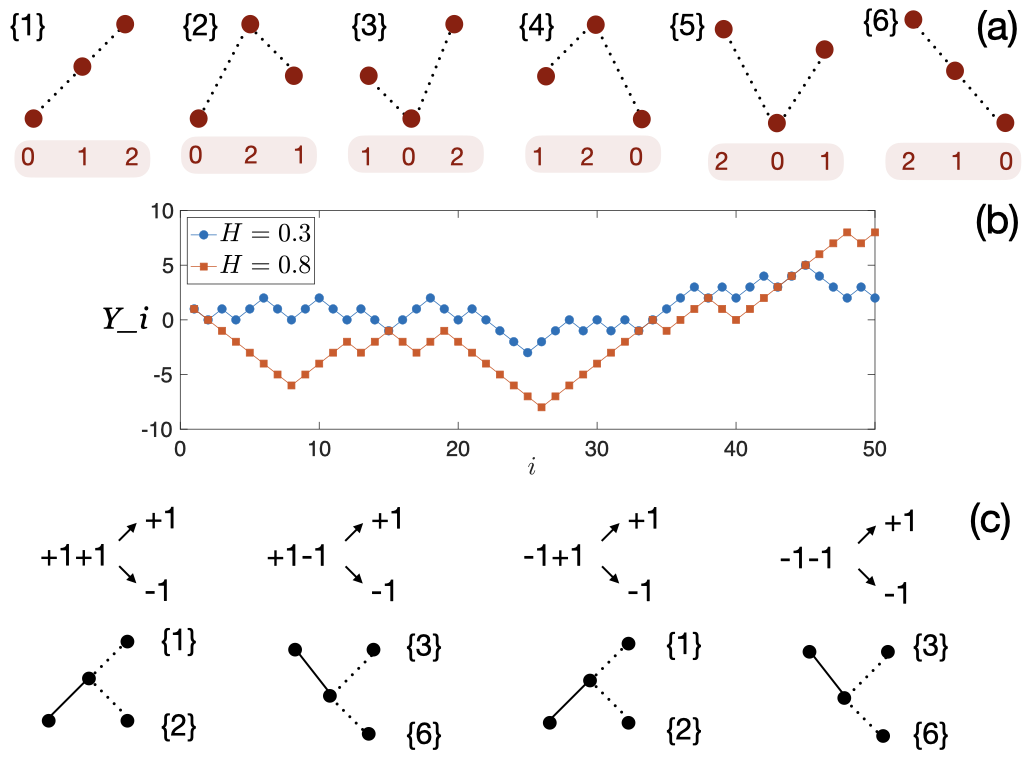}
\caption{\label{fig1} {\bf{ Ordinal Patterns and their limitation in binary walks}}. (a) Graphical representation of all six ordinal patterns for $D=3$ in lexicographic order. (b) The walk of the sign decomposition of a fGn with $H=0.3$ and 0.8, blue circles and red squares respectively. (c) The possible patterns that could appear in a walk from a binary sequence $\{-1,+1\}$ without added noise. }
\end{figure}

The harvest of an ordinal pattern $\{i\}$ firstly requires the definition of two parameters:  the pattern length $D\geqslant 2$ and the lag $\tau$ (the time separation between the data points)~\cite{bandt2002permutation}. Then, a sequence $X(t) = \{x_t ; t = 1, \dots,M\}$ can be mapped into subsets of length $D$ of consecutive ($\tau = 1$) or non-consecutive ($\tau > 1$) values, assigning to each time $t$ the $D$-dimensional vector of values at times $t$, $t + \tau , .... , t + (D-1)\tau$. Subsequently, each element of the vector is replaced by a number related to its relative ranking, {\it {i.e.}} the smallest value by zero and the largest one by $D-1$. By the ordinal pattern corresponding to the time ($t$), one therefore means the permutation $\{i\}$ of $0, 1, . . . , D - 1$, representing the relative amplitude (strength) of each element in the original vector. A graphical representation of the ordinal patterns of length $D=3$ is depicted in Fig. \ref{fig1}(a), sorted in lexicographic order. Finally, the ordinal pattern probability distribution is computed by counting the number of occurrences of each permutation, $\# (\{i\})$,  normalized by the total number of ordinal patterns $M-(D-1)\tau$. Note that the condition $M \gg D!$ must be satisfied in order to obtain a reliable statistics~\cite{bandt2002permutation}

Bandt and Shiha \cite{bandt2007order} introduced theoretical expressions for the relative frequencies of the ordinal patterns associated with different stochastic processes for patterns length 3 and 4. Particularly, for Fractional Brownian motion and $D=3$ one has
\begin{equation}\label{Eq1}
    p_{\{1\}} = \frac{1}{\pi} \arcsin(2^{H-1}), \,\,\,\,\, \forall \, \tau
\end{equation}
with $H$ being the Hurst exponent. Additionally, $p_{\{1\}}=p_{\{6\}}$ and $p_{\{2\}}=p_{\{3\}}=p_{\{4\}}=p_{\{5\}}= 1/4 -p_{\{1\}}/2$. These expressions permit us to analytically calculate all the theoretical probabilities for a given value of $H$, and vice versa. Expressions for $D=4$ can only be solved numerically. Consequently, for the present analysis, the pattern length is set to 3. This choice also helps fulfill the condition $M \gg D!$, which is particularly advantageous for real-world scenarios. Application of these theoretical expressions is useful for estimating the Hurst exponent when analyzing experimental data~\cite{olivares2016quantifying,sinn2011estimation}. Moreover, It has been shown that linear combinations of these six probabilities leads to new ways of measuring persistence, symmetry, and reversibility in time series~\cite{bandt2023statistics}. 

\section{Numerical analysis}

For illustrating, Fig. \ref{fig1}(b) shows the walk of the sign decomposition, $Y_i =\sum_{t=1}^{i} \text{sign}(x_t)$, of a synthetic fGn with $H=0.3$ and 0.8, using the method of Wood and Chan~\cite{coeurjolly2000simulation}. Naturally, the observed ordinal patterns will be directly conditioned by the low resolution of the walk, since equal consecutive values in the sequence are typically ranked according to their temporal order of appearance~\cite{bandt2002permutation}. For $D=3$, the walk of a sign decomposition generates four different patterns, as depicted in Fig. \ref{fig1}(c), showing the graphical representation for all eight possible combinations of steps of order 3. Particularly, all the triangle-shaped patterns lead to the ordinal patterns $\{2\}$ and $\{3\}$, if they point up or down, respectively. Therefore. the patterns $\{4\}$ and $\{5\}$ are somewhat ``forbidden''. For $D=4$, from each of the four possible patterns that arise when $D=3$, only two possible new patterns can emerge from each of them. The same logic applies as $D$ increases, with each materialized pattern leading to two new ones. I have numerically found that the same number of ``forbidden'' patterns exists, $\forall H \in (0,1)$, given by
\begin{equation}\label{Eq2}
    D! - 2^{D-1},
\end{equation}
as reported in the middle column in Tab. \ref{tab:forOP}.
Note that even those patterns are genuinely forbidden, they are not related to deterministic features of non linearity~\cite{amigo2006order,amigo2007true,rosso2012amigo}, but to how the methodology is implemented. To address this issue, it is necessary to add a small amount of Gaussian noise to break any ties when characterizing the walk of a sign decomposition using ordinal patterns~\cite{bandt2002permutation,zunino2017permutation}. In doing so, for $D=3$ all the $3!$ possible patterns appear. However, for $4\leq D \leq 7$, a larger noise amplitude is necessary to ensure that all patterns appear, as indicated in the right column in Tab \ref{tab:forOP}, which lists the number of unobserved patterns for sequences of up to $10^8$ data points. Nevertheless, incorporating this larger amplitude noise may affect the linear correlation within the temporal sequence of the walk. In fact, a competition between the original dynamics and the stochastic component can occur even reaching a state where the noisy component dominates and the original the dynamics can be considered as a perturbation~\cite{rosso2012amigo}. Therefore, all future analyses of ordinal patterns will be performed on the walk of the sign decomposition with a small amount of added Gaussian noise, specifically with a mean of zero and a standard deviation of $10^{-5}$

\begin{figure*}[t!]
\centering
\includegraphics[width=\textwidth]{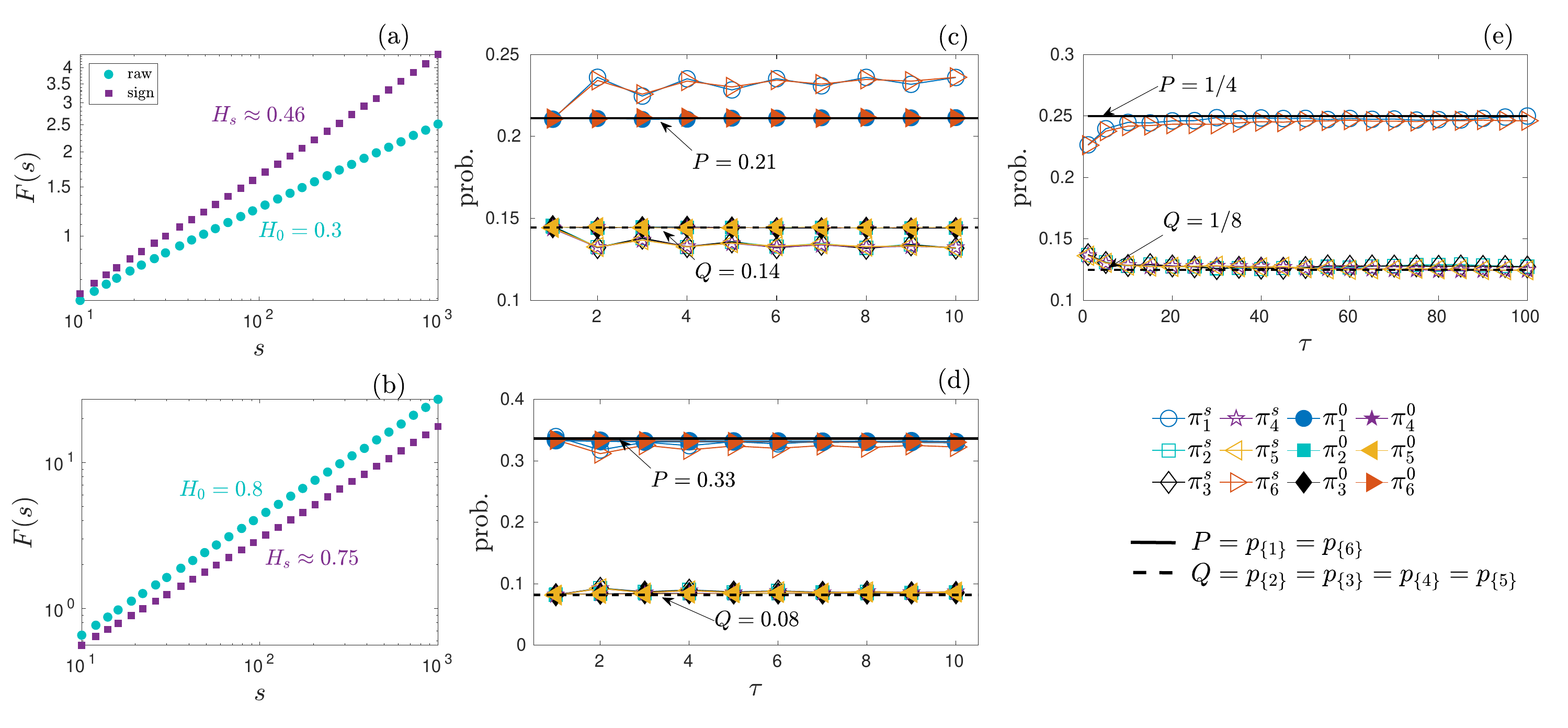}
\caption{\label{fig2} {\bf{DFA and Ordinal Patterns analysis of persistent and antipersistent sequences and their sign decomposition}}. Fluctuation function $F(s)$ versus window size $s$ for a fGn (green circles) and their sign decomposition (purple squares) for (a) $H=0.3$ and (b) $H=0.8$.  For the DFA implementation, a number of 30 window sizes $s$ equally distributed in the logarithmic scale were selected for $s \in [10, 1000]$ and a detrending polynomial of second order $m = 2$. The ordinal patterns probabilities ($D=3$) as a function of the lag for fGn (solid colored symbols) and its sign (open symbols) for (c) $H=0.3$ and (d) $H=0.8$. Finally, (e) shows the numerical convergence of the ordinal patterns probabilities with $\tau$ for the sign decomposition of fGn for $H=0.3$. Solid and dashed black lines indicate the theoretic probabilities from Eq. \ref{Eq1} for the corresponding exponent $H$. The average over 100 independent realizations is reported.}
\end{figure*}

\begin{table}[ht!]
    \centering
    \caption{List of unobserved patterns without adding noise (middle column) and with small amount of added Gaussian noise of zero mean and standard deviation of $10^{-5}$ (right column) for $D \in[3,7]$, $M = 10^8$ and $\tau=1$. These results are valid $\forall H \in (0,1)$.}
    \label{tab:forOP}
    \begin{tabular}{|c|c|c|}
 \hline
$D$ & unobserved OP & unobserved OP \\
 & (without noise) & (with noise)  \\
 \hline\hline
 3 & 2 & 0 \\ \hline
  4 & 16 & 6 \\ \hline
   5 & 104 & 50 \\ \hline
    6 & 688 & 446 \\ \hline
     7 & 4976 & 3726 \\ \hline
    \end{tabular}
\end{table}
\newpage
Now, I contrast the characterization of the walk of the sign decomposition of synthetic persistent and antipersistent fGn of $M = 10^4$ data points, by using both DFA\footnote{Briefly, DFA studies
the integrated fluctuations of a time series by systematically eliminating mth-degree polynomial trends over windows of size $s$. A fluctuation function is calculated $F(s)$ which scales as $s^H$
for correlated data. Then, a linear fit in loglog-scale over a certain window range allows to estimate the Hurst exponent~\cite{peng1994mosaic,kantelhardt2001detecting}. For its implementation, I have used the MATLAB function provided in Ref.~\cite{ihlen2012introduction}}. and the ordinal patterns probabilities, computed by the MATLAB function provided in Ref.~\cite{parlitz2012classifying}. Fig. \ref{fig2}(a) and (b) show the DFA analysis for both the raw fGn and its sign decomposition for $H=0.3$ and 0.8, respectively. It is observed that the antipersistency is not captured by any of the temporal scales accessible for the DFA methodology, i.e. $s \in [10 ,M/10]$, when analysing the sign series, contrary for the case of the persistent series---see purple squares in Fig. \ref{fig2}(b). The evolution of the ordinal patterns probabilities as a function of the lag is shown in Fig. \ref{fig2}(c) and (d) for $H=0.3$ and 0.8, respectively. On the one hand, independently of the value of $H$, the probabilities of the original fluctuations are time scale-invariant, as expected for being a self-similar process~\cite{olivares2016quantifying,bandt2020order,zunino2022permutation}. On the other hand, for the sign series, the anti-correlations are well characterized by the probabilities values only when $\tau=1$. As the lag increases, the probabilities fluctuate and converge to values that are nearly equivalent to those characterizing a random walk; $p_{\{1\}}=p_{\{6\}}=1/4$ and $p_{\{2\}}=p_{\{3\}}=p_{\{4\}}=p_{\{5\}}= 1/8$---see Fig. \ref{fig2}(e)---which is consistent with the results obtained using DFA (see Fig. \ref{fig2}(a)). 
Lastly, for $H=0.8$, the sign decomposition follows a similar evolution as the original fluctuations. According to these findings, it can be concluded that although the ordinal patterns probabilities provide a similar characterization as DFA of the dynamical changes across temporal scales (varying $\tau$), they are able to access to the high-frequency correlations when $\tau=1$, since the lag physically corresponds to multiples of the sampling time of the data~\cite{zunino2012distinguishing}. This is impossible to access with DFA due to the limitation of the minimum window size being equal to 10 data points~\cite{kantelhardt2001detecting}, therefore, high-frequency temporal information is, in a way, lost. This was empirically observed in the ordinal analysis of Human gait records~\cite{olivares2016quantifying,olivares2020multiscale}. Consequently, by setting $\tau=1$, the dynamical information of the antipersistency is successfully achieved. These results are a numerical benchmark indicating that the walk for a sign decomposition follows the same ordinal pattern probability values because it inherits the same linear correlations for lags equal to one for $H \in (0,0.5)$.

\section{Experimental applications}

In this section, ordinal pattern characterization is used to analyze experimental fluctuations and quantify the presence of anti-correlations in physiological and environmental measurements. Furthermore, these findings are contrasted with the results obtained using DFA. 


\subsection{Southern Oscillation Index fluctuations}

\begin{figure}[]
\centering
\includegraphics[width=\textwidth]{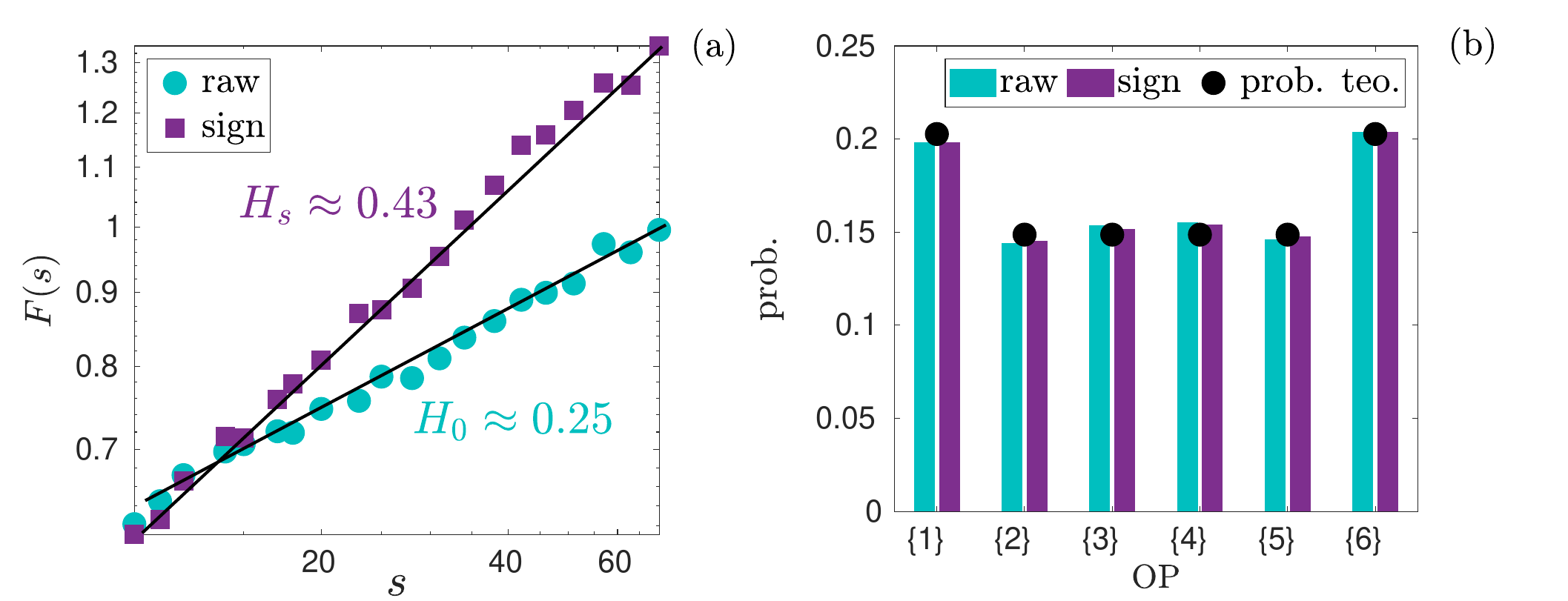}
\caption{\label{fig3}{\bf{ Capturing anticorrelations from the sign of SOI fluctuations}}. (a) Fluctuation function $F(s)$ versus window size $s$ for the raw SOI data (green circles) and its sign decomposition (purple squares). A number of 20 window sizes $s$ equally distributed in the logarithmic scale were selected for $s \in [10, 70]$ and a detrending polynomial of second order $m = 2$. (b) OP probabilities ($D=3$ and $\tau=1$) for the raw SOI data and its sign decomposition. Solid black circles indicate the theoretic probability from Eq. \ref{Eq1} with $H=0.25$. The average over the 5 sequences is depicted. }
\end{figure}

The normalized difference between the observed sea level pressure between Darwin and Tahiti is used to define the so-called Southern Oscillation Index (SOI). It has been found that SOI fluctuations can be characterized as antipersistent with an exponent $H=0.25$, for a time frame of approximately 4 months to 6 years~\cite{ausloos2001power}. The data set consists of 1620 data points measured in the years 1866-2000 ( available at https://www.cpc.ncep.noaa.gov/data/indices/). For the ordinal patterns probability calculation, five sequences of length 323 data points were generated by re-sampling the original monthly data at 5 months interval, then averaged probabilities are reported. In this manner, it is ensured that the ordinal patterns capture clean correlations and avoid dynamical transitions between different scaling~\cite{olivares2020multiscale}.  Figure \ref{fig3}(a) shows the DFA results for the original data and its sign decomposition. The DFA characterization of the sign sequence confirms the loss of scaling describing the anti-correlations, as expected, However, Fig. \ref{fig3}(b) shows that the probabilities of ordinal patterns provide evidence that the scaling is indeed inherited by the series of signs for $\tau=1$. Furthermore,  note that the empirical probabilities are in accordance with the theoretical values calculated using Eq. \ref{Eq1} when evaluating with an exponent $H=0.25$.

\begin{figure}[t!]
\centering
\includegraphics[width=\textwidth]{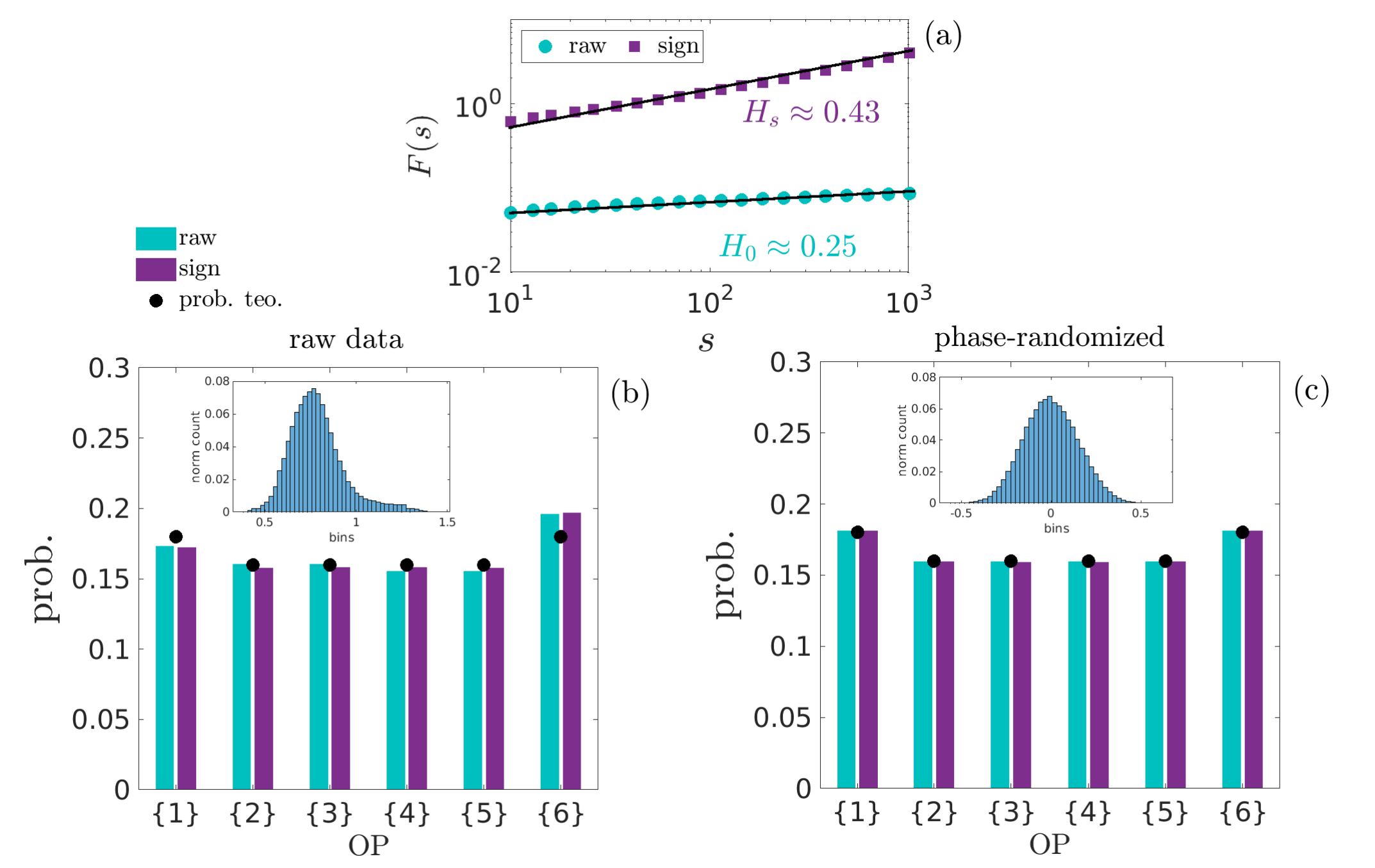}
\caption{\label{fig4} {\bf{ Capturing anticorrelations from the sign of non-Gaussian physiological fluctuations}} (a) fluctuation function $F(s)$ versus window size $s$ for the time interval sequence (green circles) and its sign decomposition (purple squares). A number of 20 window sizes $s$ equally distributed in the logarithmic scale were selected for $s \in [10, 1000]$ and a detrending polynomial of second order $m = 2$. (b) OP probabilities ($D=3$ and $\tau=1$) for the raw time intervals and its sign decomposition. Inset plot shows the histogram of the raw data of one random patient. And (c) OP probabilities obtained from the phase randomized original sequences and its sign decomposition. Inset plot shows the histogram of the phase-randomized data from the same patient. Solid black circles indicate the theoretic probability from Eq. \ref{Eq1} with $H=0.1$. The average over the 18 patients is reported}
\end{figure}

\subsection{Physiological data}
It is well known that sequential time intervals between consecutive beats of long-term ECG recordings of subjects in normal sinus rhythm are characterized by long-term anti-correlations~\cite{peng1993long,olivares2016quantifying,ashkenazy2001magnitude}. Here, I consider sequences of time intervals between consecutive beats from 18 patients (13 women and 5 men) with lengths in a range between 75,106 and 115,911 data points~\cite{goldberger2000physiobank} (Data are available at http://physionet.org/physiobank/database/nsrdb). It has been shown that this data set is characterized by a mean exponent $H=0.1$ starting from about 14 intervals onward~\cite{olivares2016quantifying}. Same as before, a decomposition into 15 sequences was generated by re-sampling every 15 time intervals, and the average of OP probabilities is then calculated for each patient. The DFA results for the original time intervals recordings and their sign decomposition are shown in Fig. \ref{fig4}(a). As expected, the scaling exhibits an exponent of around 0.1, indicating antipersistent dynamics. This behavior is not well captured by a DFA analysis when examining the sign sequences, resulting in a mean exponent of $H\sim0.43$---see purple squares in Fig. \ref{fig4}(a). The ordinal patterns probabilities, on the other hand, capture approximately the antipersistency for both the raw sequence and its sign series, as shown in Fig. \ref{fig4}(b). Yet, a noticeable difference between the first $\{1\}$ and the sixth  $\{6\}$ patterns is observed which originates from the non-Gaussian nature of the original fluctuations (non-symmetric), as evidenced in the inset plot in Fig. \ref{fig4}(b). Consequently, a surrogate analysis is conducted to obtain Gaussian distributed data with the same linear correlation, using a phase randomized procedure~\cite{theiler1992testing}. These results are shown in Fig. \ref{fig4}(c), where the experimental probabilities match perfectly with the theoretical prediction, as a consequence of the Gaussianity of the data (see inset plot).

\section{Conclusions}

The effectiveness of ordinal pattern probabilities in quantifying antipersistency in short binary sequences has been explored. Traditional methodologies such as Detrended Fluctuation Analysis and Fluctuation analysis (not discussed here, please see Ref.~\cite{carpena2017spurious}) have been found ineffective in capturing anti-correlations in binary sequences. The empirical evidence, supported by analytical expressions and numerical simulations, shows that ordinal patterns probabilities can quantify between persistent and antipersistent dynamics from totally different phenomena, including physiology and environmental sciences. Both applications have demonstrated the reliability of this approach even with non-Gaussian distributed data and very short sequence lengths. The results underscore the importance of considering the lag parameter in the analysis, as it enables the detection of high-frequency correlations often missed by DFA due to the requirement of a minimum window size of approximately 10 data points.

The dynamics of many complex systems are often studied by analyzing temporal fluctuations. These fluctuations can be represented as random walks of their sign decomposition, and the ordinal patterns probabilities offer a simple, fast, and robust against observational noise~\cite{olivares2016quantifying} approach to reveal linear correlations in the corresponding direction of the fluctuations, particularly antipersistency. I encourage researchers to incorporate the use of ordinal pattern probabilities in their analyses, as they have the ability to effectively capture the temporal structures of time series. For example, when dealing with a significant amount of data, for a given sampling rate, researchers can use Eq. (1) with $\tau=1$ to systematize the estimation of the Hurst exponent.

\section{Acknowledgements}
This work was partially supported by the Mar\'ia de Maeztu project CEX2021-001164-M funded by the  MICIU/AEI/10.13039/501100011033. The author especially thanks Massimiliano Zanin for
helpful discussions.

 \bibliographystyle{elsarticle-num} 
 \bibliography{Drfat0}

\end{document}